\documentclass[10pt,letterpaper]{article}
\usepackage[utf8]{inputenc}
\usepackage[T1]{fontenc}
\usepackage[margin=1.04in,includefoot,footskip=20pt]{geometry}
\usepackage{mathptmx,amsmath,graphicx,booktabs,array,tabularx,microtype,float}
\usepackage[font=normalsize,labelfont=bf,labelsep=period,skip=6pt]{caption}
\usepackage[numbers]{natbib}
\usepackage[hidelinks]{hyperref}
\input{glyphtounicode}
\hypersetup{pdftitle={WaveletECO: A Closed-Loop Physical ECO Platform and a Specialized Local Language Model},pdfauthor={Guoxiang Xu, Guozhen Ji, Zijian Luo, Zhengrui Chen, Qi Sun, Cheng Zhuo}}
\renewcommand{\arraystretch}{1.15}
\makeatletter
\renewcommand{\section}{\@startsection{section}{1}{\z@}{10pt plus 2pt minus 2pt}{5pt plus 1pt}{\normalfont\fontsize{12}{14}\selectfont\bfseries}}
\renewcommand{\subsection}{\@startsection{subsection}{2}{\z@}{8pt plus 2pt minus 1pt}{4pt plus 1pt}{\normalfont\fontsize{11}{13}\selectfont\bfseries}}
\makeatother
\renewenvironment{abstract}{\par\noindent\textbf{Abstract.}\ }{\par\vspace{6pt}}


\makeatletter
\renewcommand{\@maketitle}{%
  \begin{center}%
    {\fontsize{14}{17}\selectfont\bfseries\@title\par}%
    \vspace{10pt}%
    {\normalsize\begin{tabular}[t]{c}\@author\end{tabular}\par}%
  \end{center}%
}
\makeatother
\title{WaveletECO: A Closed-Loop Physical ECO Platform\\and a Specialized Local Language Model}
\author{\normalsize
Guoxiang Xu\textsuperscript{1,*}, Guozhen Ji\textsuperscript{1,*}, Zijian Luo\textsuperscript{1},\\
\normalsize Zhengrui Chen\textsuperscript{1,2}, Qi Sun\textsuperscript{1,2,\textdagger}, Cheng Zhuo\textsuperscript{1,\textdagger}\\[4pt]
\normalsize\textsuperscript{1}Zhejiang University \qquad \textsuperscript{2}ChipFlux\\
\normalsize qisunchn@zju.edu.cn \qquad czhuo@zju.edu.cn\\
\normalsize\textsuperscript{*}Equal contribution. \qquad \textsuperscript{\textdagger}Corresponding authors.}
\date{}
\begin{document}
\maketitle
\pagestyle{plain}\raggedbottom

\begin{abstract}
Engineering change order (ECO) is an important step in repairing timing and electrical violations during the late stages of chip design. Existing Agentic EDA methods primarily focus on tool invocation, with less attention to model decision quality and targeted training. A central challenge in ECO is multi-round decision-making: the model must use the results of each round to determine the next repair action. We propose WaveletECO, which integrates a closed-loop execution platform with large language models to enable agents to execute ECO decisions effectively. We also train a local 9B model through supervised fine-tuning and CPO-SimPO using execution demonstrations and decision-preference data, enabling ECO decision-making with a locally deployed model. Across 594 evaluation runs on 22 designs, WaveletECO-Policy (BF16) and (INT8) score 79.63 and 79.65, respectively, compared with GPT-6 Astra's 77.44. The estimated inference cost of INT8 is about 1/147 of GPT-6 Astra's. These results show that specialized model training supports effective, low-cost multi-round ECO repair, with repair quality retained under INT8 quantization.
\end{abstract}
\section{Introduction and Related Work}
Physical ECO~\citep{kahng2011vlsi}\citep{sapatnekar2004timing} repeatedly repairs timing and electrical violations under area constraints, requiring model decisions to be executed, verified, and retained across rounds.

Recent work brings language models into this process. ChipNeMo~\citep{liu2023chipnemo} and ChatEDA~\citep{he2024chateda} explore model adaptation for chip design and EDA, while RSR~\citep{ouyang2026rsr} and AgenticECO~\citep{ren2026agenticeco} provide ECO agent workflows. These ECO frameworks establish execution workflows but do not report dedicated training for repair decisions (Table~\ref{tab:related}).

WaveletECO trains a local 9B model on ECO demonstrations and action preferences, comparing commercial, trained, and quantized models under shared execution rules.

\begin{table}[H]
\centering
\caption{ECO methods: models, training, and task scope.}
\label{tab:related}
\begingroup\normalsize
\setlength{\tabcolsep}{7pt}\renewcommand{\arraystretch}{1.1}
\begin{tabularx}{\columnwidth}{@{}l@{\hspace{28pt}}>{\raggedright\arraybackslash}p{\dimexpr0.34\columnwidth-6pt\relax}@{\hspace{6pt}}>{\raggedright\arraybackslash}X@{}}
\toprule
Method & Model / training & Task / scale\\\midrule
\begin{tabular}[t]{@{}l@{}}IR-aware RL~\citep{jiang2024iraware}\\  (2024)\end{tabular} & R-GCN;\newline DQN training & Gate sizing: timing/power.\newline 7 designs, 45\,nm; up to 26k gates.\\\midrule
\begin{tabular}[t]{@{}l@{}}RL-CTD~\citep{xu2026rlctd}\\  (2026)\end{tabular} & Regional net;\newline RL training & Regional ECO: timing/DRC.\newline 16 designs, 14\,nm; up to 232k gates.\\\midrule
\begin{tabular}[t]{@{}l@{}}RSR~\citep{ouyang2026rsr}\\  (2026)\end{tabular} & GPT-5-mini; RAG + reflection.\newline No ECO fine-tuning reported. & Timing/power/area.\newline 8 designs, ASAP7; up to 33k\,$\mu$m$^2$.\\\midrule
\begin{tabular}[t]{@{}l@{}}AgenticECO~\citep{ren2026agenticeco}\\  (2026)\end{tabular} & Opus 4.8 / GPT-5.6;\newline frozen LLMs + agents. & 3D spacing DRC.\newline 9 cases, ASAP7; up to 1.2k nets.\\\midrule
\begin{tabular}[t]{@{}l@{}}\textbf{WaveletECO}\\ \textbf{(Ours)}\end{tabular} & \textbf{Qwen3.5-9B, local;}\newline \textbf{SFT + CPO-SimPO.} & \textbf{Setup / hold / electrical DRV.}\newline \textbf{22 designs; up to 202k cells.}\newline Five ECO rounds; three repeats per model/design.\\\bottomrule
\end{tabularx}
\par\vspace{4pt}\raggedright
Sizes are rounded maxima in source-specific units. RSR also tests GPT-4o-mini and open-weight DeepSeek-V3.2.\par
\endgroup
\end{table}
\section{Workflow and Model Training}
\begin{figure}[H]
\centering
\includegraphics[width=\textwidth]{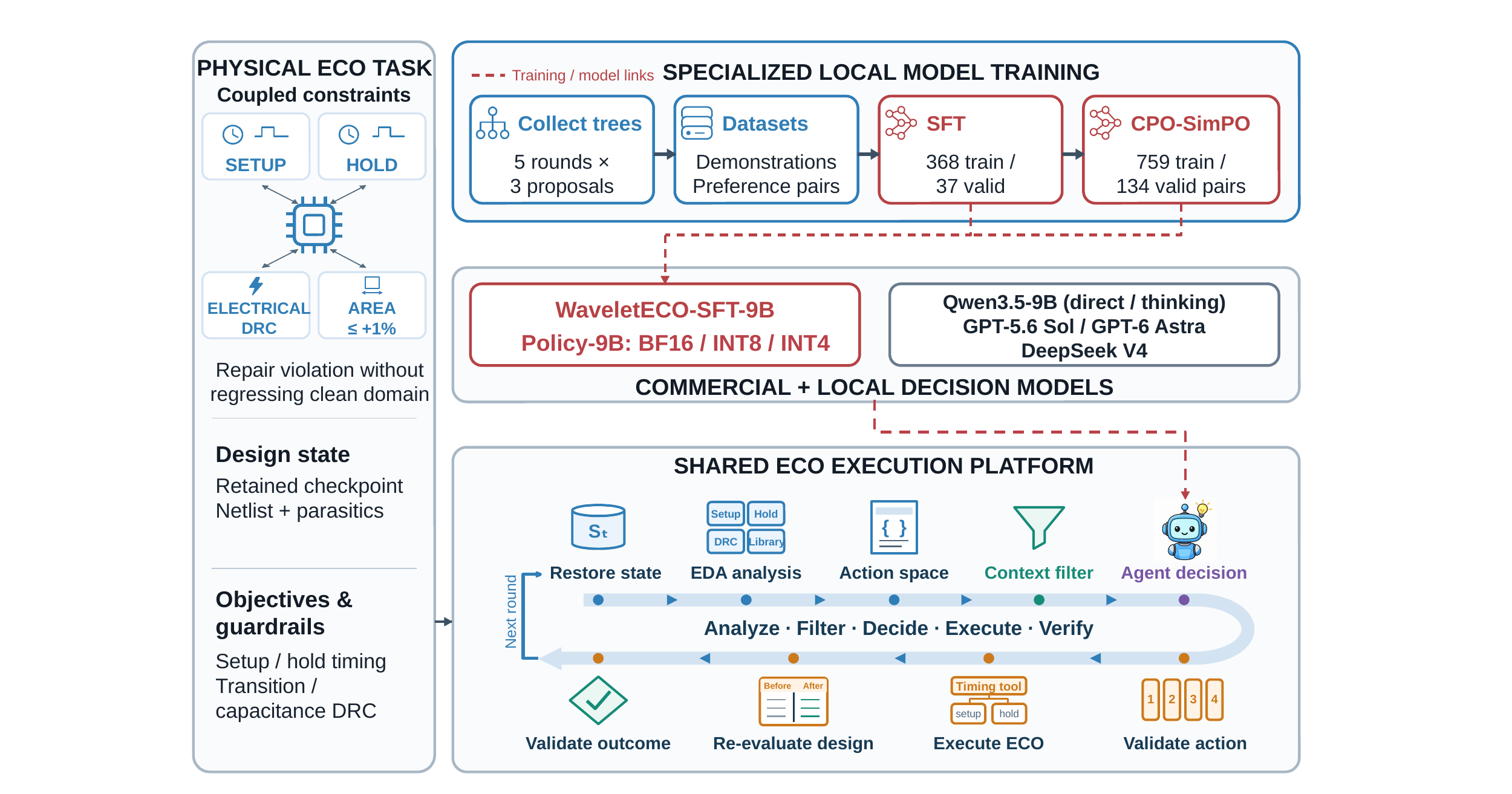}
\captionsetup{width=\textwidth}
\caption{WaveletECO combines a shared ECO workflow with training of a local repair model. Commercial and local models use the same action checks, physical measurements, and rollback mechanism.}\label{fig:overview}
\end{figure}

Each round analyzes timing, electrical, and library information (Figure~\ref{fig:overview}). The model selects a repair and its targets; the shared workflow validates, executes, and measures it, retaining accepted changes and rolling back rejected ones. Commercial and local models use the same interface.

Data collection tests three proposals per state for up to five rounds. Training and validation data for SFT and CPO-SimPO~\citep{xu2024contrastive}\citep{meng2024simpo} are split by physical design state. Offline preference labels compare same-state actions by execution success or recorded branch quality. Each response specifies one repair action. TorchAO~\citep{torchao2026} quantizes weights to INT8/INT4 with BF16 activations.

\section{Experimental Results}
\subsection{Protocol and Repair Quality}
We evaluate nine policy variants on 22 designs, with up to five repair rounds per run. Three repeats per model and design yield 594 runs. The benchmark contains 13 designs used for local adaptation or model selection and nine held-out designs. All variants use the same EDA backend, one candidate per round, and a 1\% area-growth limit.

Let $b_{d,k}$ and $x$ denote the initial and terminal values of metric $k$ for design $d$. Violation magnitude is $v_k(x)=\max(0,-x)$ for WNS/TNS and $v_k(x)=x$ for violation counts. Each metric receives
\par\begingroup
\setlength{\abovedisplayskip}{7pt}
\setlength{\belowdisplayskip}{7pt}
\setlength{\abovedisplayshortskip}{4pt}
\setlength{\belowdisplayshortskip}{4pt}
\begin{equation}
q_{d,k}(x)=
\begin{cases}
100\,\operatorname{clip}\!\left(1-\dfrac{v_k(x)}{v_k(b_{d,k})},0,1\right),
& v_k(b_{d,k})>0,\\[4pt]
100, & v_k(b_{d,k})=0,\ v_k(x)=0,\\
0, & v_k(b_{d,k})=0,\ v_k(x)>0.
\end{cases}
\label{eq:metric}
\end{equation}
\nopagebreak[2]
The terminal ECO score equally weights setup, hold, and electrical repair:
\begin{equation}
S_{d,r}=\frac{1}{3}\left[
\frac{q_{W_S}+q_{T_S}+q_{N_S}}{3}
+\frac{q_{W_H}+q_{T_H}+q_{N_H}}{3}+q_D\right].
\label{eq:score}
\end{equation}
Here, $W,T,N$ denote WNS, TNS, and violation count; subscripts $S/H$ indicate setup/hold, and $D$ denotes the deduplicated electrical violation count. Scores use unrounded measurements and are averaged over three repeats, then equally across designs.
\par\endgroup

\begin{table}[H]
\centering
\caption{Repair quality and estimated inference cost on 22 designs.}
\label{tab:results}
\begingroup
\normalsize
\setlength{\tabcolsep}{3pt}
\renewcommand{\arraystretch}{1.12}
\begin{tabular}{@{}lrrrrrr@{}}
\toprule
& \multicolumn{3}{c}{ECO score (mean $\pm$ SD)} & \multicolumn{3}{c}{All-22 cost / efficiency}\\
\cmidrule(lr){2-4}\cmidrule(l){5-7}
Model & Seen & Unseen & All (22) & \shortstack{USD/task\\$\bar C_m$} & \shortstack{Score/USD\\$E_m$} & \shortstack{Relative\\$R_m$}\\
\midrule
Initial state & $51.28 \pm 0.00$ & $62.96 \pm 0.00$ & $56.06 \pm 0.00$ & --- & --- & ---\\
\addlinespace[2pt]
\begin{tabular}[c]{@{}l@{}}Qwen3.5-9B\\(direct)$^\dagger$\end{tabular} & $66.41 \pm 0.00$ & $83.00 \pm 0.00$ & $73.20 \pm 0.00$ & 0.007 & 9,784.1 & $41.93\times$\\
\begin{tabular}[c]{@{}l@{}}Qwen3.5-9B\\(thinking)$^\dagger$\end{tabular} & $66.71 \pm 0.29$ & $80.08 \pm 0.64$ & $72.18 \pm 0.43$ & 0.088 & 820.5 & $3.52\times$\\
WaveletECO-SFT$^\dagger$ & $70.92 \pm 0.00$ & $86.49 \pm 0.00$ & $77.29 \pm 0.00$ & 0.005 & 14,339.0 & $61.45\times$\\
\begin{tabular}[c]{@{}l@{}}WaveletECO-Policy\\(BF16)\end{tabular} & $74.44 \pm 0.34$ & \textbf{87.13 $\pm$ 0.08} & $79.63 \pm 0.17$ & 0.008 & 10,135.8 & $43.44\times$\\
\begin{tabular}[c]{@{}l@{}}\textbf{WaveletECO-Policy}\\\textbf{(INT8)}\end{tabular} & \textbf{74.84 $\pm$ 0.00} & \textbf{86.58 $\pm$ 0.75} & \textbf{79.65 $\pm$ 0.31} & \textbf{0.002} & \textbf{35,208.6} & $\mathbf{150.89}\times$\\
\begin{tabular}[c]{@{}l@{}}WaveletECO-Policy\\(INT4)\end{tabular} & $73.43 \pm 0.02$ & $86.24 \pm 0.00$ & $78.67 \pm 0.01$ & \textbf{0.001} & \textbf{53,085.4} & $\mathbf{227.50}\times$\\
GPT-5.6 Sol & $70.36 \pm 0.19$ & $84.52 \pm 0.73$ & $76.15 \pm 0.36$ & 0.161 & 471.7 & $2.02\times$\\
GPT-6 Astra & $71.84 \pm 0.11$ & $85.54 \pm 0.15$ & $77.44 \pm 0.00$ & 0.332 & 233.3 & $1.00\times$\\
DeepSeek V4 & $68.85 \pm 1.06$ & $82.41 \pm 0.29$ & $74.40 \pm 0.65$ & 0.011 & 7,055.9 & $30.24\times$\\
\bottomrule
\end{tabular}
\endgroup
\par\vspace{3pt}\noindent\begin{minipage}{\textwidth}\normalsize
\textbf{Notes:} Scores report mean $\pm$ population SD over three repeats, with designs equally weighted within each repeat. SD denotes standard deviation. Seen comprises 13 designs used for adaptation or model selection; Unseen comprises 9 held-out designs. Cost and efficiency use all 22 designs. DeepSeek costs use the arithmetic mean of peak and off-peak prices. $\dagger$ Costs estimated using WaveletECO-Policy BF16 throughput. Bold marks column-best values and additionally highlights the INT8 row.
\end{minipage}
\end{table}

Table~\ref{tab:results} shows that WaveletECO-Policy (BF16) scores 79.63 across all 22 designs, compared with 77.29 for WaveletECO-SFT and 73.20 for Qwen3.5-9B (direct). Preference training improves the mean score over SFT by 2.34 points. GPT-6 Astra scores 77.44 under the same execution protocol. The INT8 and INT4 variants score 79.65 and 78.67, respectively, retaining most of the BF16 repair quality.

On the nine held-out designs, WaveletECO-Policy (BF16) scores 87.13, compared with 86.49 for WaveletECO-SFT and 85.54 for GPT-6 Astra. These results support transfer to unseen designs within this benchmark.

\subsection{Inference Cost and Cost Efficiency}
\begingroup
\setlength{\abovedisplayskip}{5pt}
\setlength{\belowdisplayskip}{5pt}
\setlength{\abovedisplayshortskip}{3pt}
\setlength{\belowdisplayshortskip}{3pt}
Mean cost over $K=66$ tasks includes every request $j$, including corrections and rejected actions:
\begin{equation}
\bar C_m=\frac{1}{K}
\begin{cases}
\displaystyle\sum_j\sum_b \frac{p_{m,b}N_{m,j,b}}{10^6}, & \text{API},\\[6pt]
\displaystyle\frac{h_m}{3600}\sum_j\left(\frac{N^{\mathrm{in}}_{m,j}}{v^{\mathrm{pre}}_{m,j}}+
\frac{(N^{\mathrm{out}}_{m,j}-1)_+}{v^{\mathrm{dec}}_{m,j}}\right), & \text{local}.
\end{cases}
\label{eq:inference_cost}
\end{equation}
Here $j$ indexes requests, $b$ API billing categories, and $N$ token counts; $p$ is USD/million tokens, $h$ USD/hour, and $v$ tokens/s. Prefill includes the first output token, with $(x)_+=\max(x,0)$. For mean ECO score $\bar S_m$, we define
\begin{equation}
E_m=\frac{\bar S_m}{\bar C_m},\qquad
R_m=\frac{E_m}{E_{\mathrm{GPT\text{-}6}}}.
\label{eq:cost_efficiency}
\end{equation}
Here $E_m$ is score per dollar and $R_m$ uses GPT-6 as $1\times$.
\par\endgroup

API costs use the fixed September 19, 2026 reference prices, including cache discounts and billed reasoning: high for GPT-5.6/GPT-6~\citep{openai2026cost}, none for DeepSeek V4-Flash-0731~\citep{deepseek2026cost}; DeepSeek costs use the arithmetic mean of off-peak and peak prices. Local single-GPU vLLM costs use token counts and independent throughput measurements on RTX 6000D (BF16) and RTX 4090 (INT8/INT4), excluding EDA and idle time. Rental proxies are CNY 6.98/hour (6000D) and 1.88/hour (4090)~\citep{autodl2026cost}, at 6.6996 CNY/USD.

\begin{figure}[!t]
\centering
\includegraphics[width=0.82\textwidth]{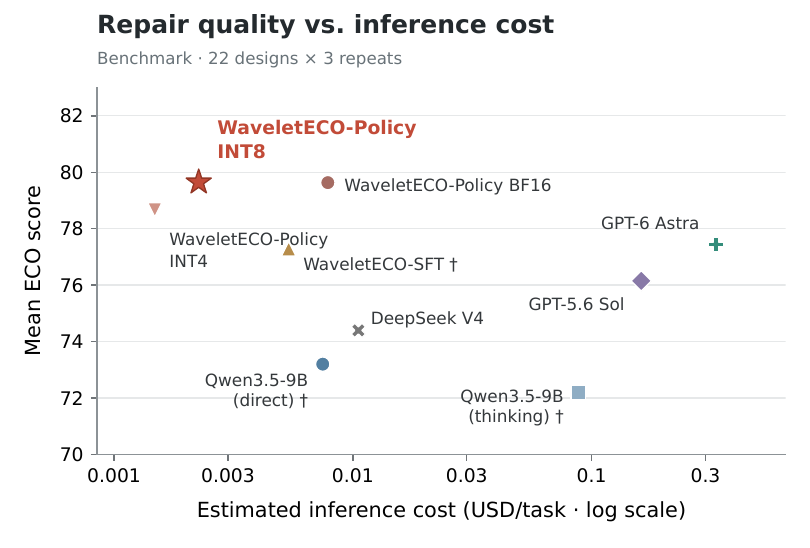}
\captionsetup{width=\textwidth}
\caption{Repair quality versus estimated inference cost on 22 designs (66 tasks/model). Star: WaveletECO-Policy INT8, our highest-scoring local variant. DeepSeek: arithmetic mean of off-peak and peak prices; $\dagger$: BF16 throughput proxies.}\label{fig:cost_score}
\end{figure}

\begin{samepage}
WaveletECO-Policy INT8 scores 79.65 at USD 0.00226/task versus GPT-6's 77.44 at USD 0.33188/task (Figure~\ref{fig:cost_score}), yielding $150.89\times$ the score per dollar. INT4 scores 78.67 at USD 0.00148/task, trading a small score decrease for lower cost. Repair scores are from the original evaluations; the deployments used for cost estimation have not been re-evaluated for repair quality.
\par\end{samepage}

\begin{samepage}
\section{Conclusion}
We present WaveletECO, which supports multi-round ECO decision-making through a closed-loop execution platform and improves the decision quality of a local 9B model through supervised fine-tuning and preference training. Experiments show that WaveletECO achieves higher overall repair scores than GPT-6 Astra across 22 designs. In particular, the INT8 variant achieves approximately 103\% of GPT-6 Astra's repair score at an estimated inference cost of about 1/147 of GPT-6 Astra's. These results show that combining closed-loop execution with specialized model training enables locally deployed models to perform effective multi-round ECO repair at low cost. However, generalization to larger-scale and more diverse designs and technology nodes requires further validation.
\label{page:body_end}
\par\end{samepage}
\newpage
\bibliographystyle{IEEEtran}
\bibliography{references}
\end{document}